\documentclass[12pt,a4paper]{article}

\usepackage{amsmath,amssymb,amsfonts}
\usepackage[dvips]{graphicx}
\usepackage{epsfig}
\usepackage{cite}
\usepackage{amsmath,amssymb,amsfonts,graphicx}
\usepackage{epsfig}

\newcommand{\beq}{\begin{eqnarray}}
\newcommand{\eeq}{\end{eqnarray}}
\newcommand{\bea}{\begin{eqnarray}}
\newcommand{\eea}{\end{eqnarray}}

\newcommand{\bec}{\begin{center}}
\newcommand{\eec}{\end{center}}

\def\R{\mathbb{R}}
\def\CP{\mathbb{CP}}
\def\Z{\mathbb{Z}}
\def\tc{\tilde{c}}
\def\ts{\tilde{s}}
\def\tlambda{\tilde{\lambda}}

\numberwithin{equation}{section}

\begin{document}

\title{
\vskip 20pt
\bf{\Large Clustering and decomposition for non BPS solutions of the $\CP^{N-1}$ models}
\vskip 30pt}
\author{
S. Bolognesi and W. Zakrzewski \\[30pt]
{\em \normalsize  
Department of Mathematical Sciences,}\\[0pt] 
{\em \normalsize Durham University, Durham DH1 3LE, U.K.}\\ 
{\normalsize Email: s.bolognesi@durham.ac.uk,   w.j.zakrzewski@durham.ac.uk} 
}
\vskip 10pt
\date{October 2013}
\maketitle
\vskip 20pt
\begin{abstract}
We look at solutions (both BPS and non-BPS) of the $\CP^{N-1}$ model on $\R \times S^1$ (with twisted boundary conditions), in particular by using a conformal mapping technique, and we show how to interpret these solutions by decomposing  them into expressions describing  constituent solitons.  We point out the problems that may arise (for non-BPS solutions) when one naively looks at the clustering properties of these solutions.
This could lead to misunderstandings when studying extrapolations between small and large compactification radii. 
\end{abstract}
\newpage

\section{Introduction}

The $\CP^{N-1}$ non-linear sigma model on the Euclidean plane $\R^2$ has holomorphic and anti-holomorphic solutions which saturate the BPS bound and minimize the action in any given topological sector \cite{D'Adda:1978uc}.  When $N>2$ there are also  additional solutions, which are not BPS and which are only unstable saddle points of the action. 
These solutions have been extensively studied in the past on the $\R^2$ and $S^2$ backgrounds \cite{Din:1980jg,Din:1980uj,Din:1980wh,book}.  There exists a solution generating technique which allows to construct these non-BPS solutions from the holomorphic (or anti-holomorphic) ones by acting on them, several times,  with certain  operators $P_{\pm}$. A number of theorems have been rigorously established; in particular the proof that this procedure is complete, in the sense that all such solutions can be obtained by the repeated action of one of these operators on the holomorphic (or antiholomorphic)  solutions.

Some recent papers have revitalized the interest in non-BPS solutions of sigma models defined on Euclidean space  and, in particular, also on the cylinder $\R \times S^1$ \cite{Dabrowski:2013kba,Cherman:2013yfa,Basar:2013eka}.  Compactification deforms the original QFT by introducing an infra-red cutoff, thus bringing the theory to a region in which it becomes tractable as a semi-classical quantum mechanical system. 
The main goal of these approaches is to try to provide a rigorous mathematical definition of the QFT by invoking the principle of continuous connection between small and large compactification radii \cite{Dunne:2012ae,Dunne:2012zk,Argyres:2012ka}. Establishing this continuous connection is probably one of the main difficult points that have to be clarified to achieve 
the success of this program.

Holomorphic solutions of the $\CP^{N-1}$ sigma model  on   $\R \times S^1$ have been studied  in \cite{Bruckmann:2007zh,Brendel:2009mp,Harland:2009mf} and they have many properties in common  with instantons on $\R^3 \times S^1$.
In the $\CP^{N-1}$ the boundary conditions are parametrized  by $N$ real angles $\theta_I$.  The ``strictly'' periodic boundary conditions correspond to the case when all the angles are  equal.  When the angles are  maximally separated, {\it i.e.} $\theta_I = 2 \pi I / N$ we say that we have the so called ``twisted'' boundary conditions. 
Continuity between small and large compactification radii works very well in the holomorphic sector.  As the size of the compactified direction is varied, while keeping the soliton size fixed,  the action and the number of zero modes around a given solution remain constant. 
However, the properties of the solutions do change.  For example, for a generic choice of the  angles $\theta_I$,  a soliton  (topological charge  one) may split into ``partons'' which carry  a fraction of the  topological charge. 

For the twisted boundary conditions a one-soliton splits into $N$ partons which carry $1/N$'th of the topological charge and  are related by a residual $\Z_N$  symmetry.   This is a crucial effect which allows us to relate the strong-coupling effects of the original QFT defined on the plane to this modified theory in a weak-coupling regime and thus to investigate their properties in a semi-classical expansion.

The main goal of this paper is to extend these studies to a non-holomorphic case, {\it i.e.} to study in detail the non-holomorphic  solutions on $\R \times S^1$, placing particular emphasis on trying to understand  how in this case  they interpolate between small and large compactification radii.   In Section \ref{cpn} we rephrase some known results about the compactified $\CP^{N-1}$ sigma model by using a conformal mapping technique and mapping from  $\R \times S^1$ to $\R^2$.  This approach will allow us to establish some new results on the properties of non-holomorphic solutions.   In Section \ref{nonholosection} we recall some facts about the operators $P_{\pm}$  which will be very useful in order to study the properties of the non-holomorphic solutions on $\R \times S^1$.  Section \ref{example} discusses the behaviour of such non holomorphic solutions in the embedding and clustering limits and  shows that, in many aspects,  this behaviour is different from the behaviour of the holomorphic solutions.

\section{$\CP^{N-1}$ sigma model on a plane and on a cylinder}
\label{cpn}

The complex projective space $\CP^{N-1}$ model can be  parametrized by a vector with $N$ complex components $n=(n_1,\dots,n_N)$ of unit norm ${n}^{\dagger}  n   = 1$ supplemented by gauging away the overall $U(1)$ phase $n \equiv e^{i \alpha} n$. The action of such a sigma model on the Euclidean plane $(x_1,x_2)$ is 
\beq
\label{action}
S = \int d^2x \  D_{\mu}n^{\dagger}   D_{\mu} n
\eeq
where the covariant derivative is  $D_{\mu} = \partial_{\mu} - i A_{\mu}$ and the gauge field is a composite of the field $n$ itself, and is given by $ A_{\mu} = - i {n}^{\dagger}  \partial_{\mu} n$.  The Euler-Lagrange equations for the field $n$ are 
given by 
\beq
\label{elequation}
  D_{\mu}  D_{\mu}  n - (n^{\dagger}  D_{\mu} D_{\mu}n  )\,  n= 0
\eeq
together with the constraint ${n}^{\dagger}  n   = 1$.
The action is classically conformally  invariant, so any conformal transformation maps solutions of the Euler-Lagrange equations into other solutions.   We will use this property very frequently in this paper.

A cylinder can be obtained from a plane by imposing on it periodicity  in one direction, say  $x_1=x_1+L$ 
and  we can  set  $L=1$ by conformal rescaling.
The $\CP^{N-1}$ field can  be taken to be  ``almost'' periodic, that is periodic up to a unitary matrix transformation $\Omega \in U(N)$
\beq
\label{almostperiodic}
 n(x_1+1,x_2) = \Omega \ n(x_1,x_2). 
\eeq
We can diagonalize $\Omega$ and recast it into a canonical form
\beq
\label{omega}
\Omega =  {\rm diag}(1,e^{i\theta_1}, \dots , e^{i\theta_{N-1}})
\eeq
with $\theta_1 \leq \dots \leq \theta_{N-1}$.  There are, in general, $N$ fixed points of this transformation in  $\CP^{N-1}$  which  are $n \propto (0,\dots,1,\dots,0)$.

Note that, being the  classical theory scale invariant,  solutions depend only on the ratios of physical length scales; for example the ratio of the instanton size $\lambda$ and of the compactification period $L$.  The two limits, of small and large compactification periods, which are properly defined by changing $L$ while keeping the instanton scale $\lambda$ fixed, can be equivalently studied by keeping the length of the period fixed while changing $\lambda$. Hence in this paper we frequently refer to the limit $\lambda \to 0$ with $L=1$ as the de-compactification limit.

Next we parametrize the $\CP^{N-1}$ space  by using a vector $w$ with $N-1$ complex components $w_{j}$ where $j=1,\dots,N-1$ defined by
\beq
\label{nfromw}
n = \frac{1}{\sqrt{1+|w|^2}}
 \left( \begin{array}{c} 
 1 \\
w_j \end{array} \right).
\label{aa}
\eeq 
Strictly speaking, in this formulation, we need more patches in order to cover entirely the $\CP^{N-1}$ space. 
This can be done by using also $u_j=\frac{1}{w_j}$. The metric in the $w_j$ coordinates is the Fubini-Study one 
\beq
\label{metricfb}
g_{\bar{i}j} = \partial_{\bar{i}} \partial_{j} \log{(1+|w|^2)},
\eeq
where $|w|^2 = \sum_{j=1}^{N-1} |w_j|^2$. By defining, for simplicity,  $ \bar{w} \cdot v  = g_{ \bar{i} j  } \, \bar{w}_{\bar{i}} v_j  $, the action of the sigma model can be rewritten as
\beq
S = \int d^2x \    \partial_{\mu} \bar{w} \cdot  \partial_{\mu} w.
\eeq

It is convenient to introduce complex coordinates to parametrize the plane  $z=x_1+ix_2$, $\bar{z}=x_1-ix_2$ and introduce corresponding complex derivatives $\partial =( \partial_{x_1} -i \partial_{x_2})/\sqrt{2}$, $\bar{\partial}= (\partial_{x_1} + i \partial_{x_2})/\sqrt{2}$. Then  the action   becomes
\beq
\label{energycomplex}
S = \int d^2z \    \bar{\partial} w \cdot \partial \bar{w} + \partial w \cdot \bar{\partial} \bar{w} 
\eeq
Note that (\ref{energycomplex}) is already in a Bogomoln'y form.  For example, for  holomorphic solutions $\bar{\partial} w = \partial \bar{w} =0$ and so the first term vanishes while  the second term, by using (\ref{metricfb}),  becomes  a total derivative  $  \partial w \cdot \bar{\partial} \bar{w}  = \partial \bar{\partial} \log{(1+|w|^2)} $. 
 The action can then be calculated by using the divergence theorem in 2 dimensions and is given by $S=2 \pi k$ where $k$ is the maximal degree of the rational functions $w_i$ (assuming no overall factors).

\begin{figure}[h!]
\epsfxsize=7.0cm
\centerline{\epsfbox{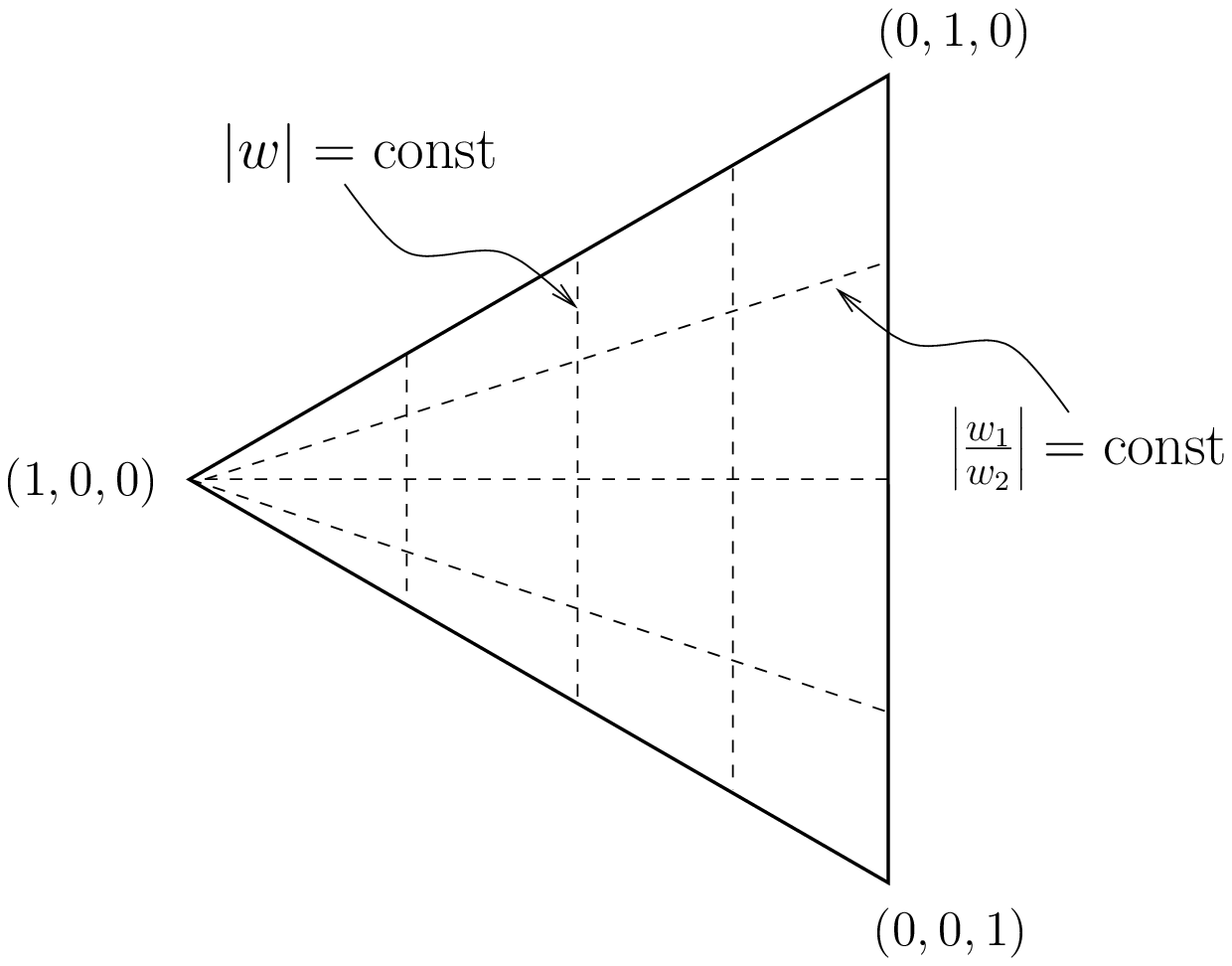}}
\caption{{\footnotesize Toric diagram for $\CP^2$.}}
\label{toric}
\end{figure}

In a toric formulation the $\CP^{N-1}$ space corresponds to a $N-1$ dimensional real torus fibred over a $N-1$ real dimensional simplex.  An example for $\CP^2$ is presented in Figure \ref{toric}. The torus shrinks to a circle on the edges of the simplex and shrinks to a point on the vertices, which can be chosen to be the fixed points of $\Omega$.

Next we consider the case of the $\CP^1$ sigma model on the cylinder \cite{Bruckmann:2007zh}.
The boundary conditions (\ref{omega}) are parametrized by one  phase $\theta$ and BPS  solutions are given by a single  holomorphic function $w(z)$ with periodicity  $w(z+1) = e^{i \theta} w(z)$. Linearity of the holomorphic solutions allows us to construct the generic one-instanton solution by summing $1/(z-m)$ poles placed on a one-dimensional lattice
\beq 
\label{infinitesum}
 w =  \sum_{m = -\infty}^{\infty} \frac{\lambda e^{i  m \theta}}{z - m}\ 
\eeq

This sum is formally divergent but it can be regularized, for example by computing its derivative first, which is convergent
\beq 
\label{infinitesumderivative}
 \frac{dw}{dz} = - \sum_{m = -\infty}^{\infty} \frac{\lambda e^{i  m \theta}}{(z - m)^2}\ ,
\eeq
and then fixing the integration constant by requiring the symmetry $w(\bar{z}) = \bar{w}(z)$ which is the one respected by every single term in (\ref{infinitesum}).
For periodic and twisted boundary conditions this sum gives respectively 
\bea
\label{solutionsone}
w = \frac{\lambda \pi}{\tan{\left(\pi z\right)}} \qquad &{\rm for}& \qquad \theta = 0\ , {\rm \ periodic}, \\
\label{twistedonesoliton}
w= \frac{\lambda \pi}{\sin{\left(\pi z\right)}} \qquad &{\rm for}& \qquad \theta = \pi\ , {\rm\  twisted}.
\eea

Note than for generic $\theta$ the value of $\CP^1$  at $x_2 = \Im{(z)} \to \pm \infty$ is not the one of the single instanton poles $1/(z-m)$, which would give $n \propto  (1,0)$.  This happens only for the special case of $\theta = \pi$ for which we have a maximal cancellation due to the alternating phases of the poles.

Let us now study in detail the twisted solution in  (\ref{twistedonesoliton}) and add a generic real parameter $a$ which corresponds to the position in the $x_2$ coordinate
\beq
\label{onesolitontwisted}
w= \frac{\lambda \pi}{\sin{\left(\pi (z-i a)\right)}}.
\eeq
We can isolate a parton inside the instanton by sending $\lambda$ to infinity while keeping  $b_1$ defined by
\beq
\label{posuno}
b_1   = a+  \frac{\log{(2  \pi \lambda)}}{\pi}
\eeq
fixed. 
This limit gives us
\beq
\label{partonone}
 \qquad w \to  -i e^{i  \pi (z-i b_1)}\qquad &{\rm for}& \qquad
\lambda \to \infty \ , \ b_1 = {\rm fixed}. 
\eeq

This limiting field corresponds to a kink in the $x_2$ direction, which interpolates between the fixed points of $\Omega$,  $n \propto (1,0)$ at $x_2 \to +\infty$  and $n\propto (0,1)$ at $x_2 \to -\infty$.  In the $x_1$ direction it is just rotating with a phase $e^{i \pi x_1}$  so it  maps the fundamental period of the  cylinder into half of the $\CP^1$ sphere  and thus it represents a parton of topological charge $1/2$.  Note that, in general, this solution is parametrized only by two moduli, the position of the kink $b_1$ and a global phase which can also be added.  

Note that we can also isolate the other parton if we define its position as
\beq
\label{posdue}
b_2   =  a - \frac{\log{(2  \pi \lambda)}}{\pi}
\eeq
and then we send $\lambda$ to infinity but this time keeping $b_2$ fixed
\beq
\label{partontwo}
 \qquad w \to  +i  e^{-i  \pi (z-ib_2)}\qquad &{\rm for}& \qquad
\lambda \to \infty \ ,\  b_2 = {\rm fixed} 
\eeq

The charge of this kink is the opposite of the previous one since  it interpolates between $n \propto (0,1)$ at $x_2 \to -\infty$ and $n \propto (1,0)$ at $x_2 \to +\infty$.
The phase rotation in the $x_1$ direction is also opposite $e^{-i \pi x_1}$ and thus the topological charge is the same as before.
So as $\log{\lambda} \gg 1$, the soliton (\ref{onesolitontwisted}) splits into two partons, each carrying half of the topological charge, and located at the positions (\ref{posuno}) and (\ref{posdue}) ({\it i.e.}  with the distance $2 \log{(2  \pi \lambda)}/\pi$ between them).   The solution becomes almost translationally invariant in the $x_1$ coordinate, as we can always perform a phase rotation on it.

\begin{figure}[h!]
\epsfxsize=10.0cm
\centerline{\epsfbox{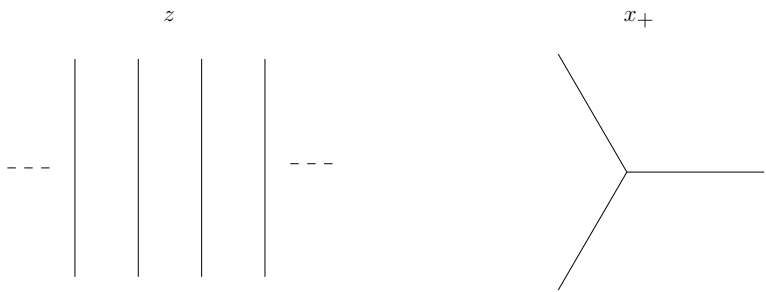}}
\caption{{\footnotesize Conformal map for $\CP^2$ with twisted boundary conditions.}}
\label{map}
\end{figure}

We now show that there is another quick route to arrive at the same result.
We take the conformal mapping
\beq
\label{conformalmapping}
x_+ = e^{-i  2 \pi z / h}     \qquad \qquad h \in \mathbb{Z},
\eeq
which maps the cylinder with identification $z \simeq z+h$ onto the plane $x_+$.  For the strictly periodic boundary conditions, {\it i.e.} for $\theta_i =0$,  we can choose $h=1$ since the function is already periodic on the fundamental period of the cylinder. 
For generic boundary conditions,  $h$ must be taken to be the smallest integer for which the matrix  $\Omega$ satisfies 
\beq
\label{omegah}
\Omega^h = {\bf 1}_{N,N}
\eeq 
Note that this is possible only if $\theta_i$ are  rational multiples  of $\pi$.
With this trick we can  map the problem onto the plane $x_+$ by imposing strict periodicity in $\arg{(x_+)}$.  For example for the twisted boundary in the $\CP^1$ case we have to choose $h=2$ so that the cylinder contains two of the fundamental periods.   For  $\CP^2$ with twisted boundary conditions we have to choose $h=3$; the corresponding map for this example is shown in Figure \ref{map}. 

Next we re-derive the previous results for the twisted boundary conditions  in the $\CP^1$ case from the $x_+$ plane perspective.  The one-instanton solution  (\ref{twistedonesoliton}) in the $x_+$ plane corresponds to two solitons located at $\pm e^{\pi a}$
{\it i.e.} $w$ is given by
\beq
\label{wfunction}
w = - 2 \pi \lambda e^{ \pi a}\left(\frac{1}{x_++e^{\pi a}} + \frac{1}{x_+ - e^{\pi a}}\right).
\eeq
In general, all  solutions in the $x_+$ plane  have the following  $\Z_2$ symmetry: 
\beq
w(-x_+) = - w(x_+).
\eeq

Note that in the $w$ plane we have a branch-cut for the inverse function of (\ref{wfunction}). This cut goes from $w=-2i\pi \lambda$ to  $w =2i\pi \lambda$. This cut is crossed each-time we move by one fundamental period of the cylinder. The widening of the cut as $\lambda \to \infty$ is another manifestation of  the fractionalization.

This can be derived from a solution on the cylinder with twisted boundary conditions and vice-versa. It is also possible to have a single instanton with the $\Z_2$ symmetry in the $x_+$ plane, but only if we fix its position to be at the origin.  In particular, the following one-soliton solutions in $x_+$ are allowed
\beq
\label{partonxplus2}
w = \frac{ e^{\pi b_1}}{x_+},  \qquad \qquad w  = - e^{\pi b_2} x_+ .
\eeq
They correspond, respectively, to the two partons (\ref{partonone}) and (\ref{partontwo}). So the partons on the $z$ cylinder correspond to one-solitons in the $x_+$ plane and fractionalization is just a consequence of the choice $h=2$ for the conformal mapping (\ref{conformalmapping}).

We can use this technique to describe partons in $\CP^2$.
The twisted boundary conditions correspond to
\bea
\Omega = {\rm diag} (1,\omega,\omega^2) \label{twistedconditioncp2}
\eea
with $\omega = e^{i 2 \pi/3}$. So to go further we have to choose  a conformal transformation (\ref{conformalmapping}) with $h=3$ in order to have strict periodicity in  $\arg{(x_+)}$. Thus we have to restrict our attentions to solutions which have the following $\Z_3$ symmetry
\beq
\label{restriction}
\left( \begin{array}{c} 
w_1(\omega x_+) \\
w_2(\omega x_+) \end{array} \right) =
\left( \begin{array}{c} 
\omega\, w_1(x_+) \\
\omega^2 w_2(x_+) \end{array} \right).
\eeq 
Thus one-soliton solutions in the $x_+$ plane with this $\Z_3$ symmetry  are given by
\bea
\left( \begin{array}{c} 
w_1 \\
w_2 \end{array} \right) = \ 
 \left(\begin{array}{c}  
 \zeta/x_+\\ 
0
\end{array}
 \right),
  \qquad 
\left( \begin{array}{c} 
w_1 \\
w_2 \end{array} \right) =
 \left( \begin{array}{c}  
0\\ 
\zeta x_+
\end{array}
 \right),
\eea
which are the direct generalizations of (\ref{partonxplus2}) and $\zeta$ is an arbitrary complex number.   Each of these solutions in the $z$ cylinder case corresponds to a parton with fractional charge $1/3$, and again this follows directly from the choice $h=3$ we have made earlier. The first parton is described by a kink that interpolates from $n \propto (1,0,0)$ to $n \propto (0,1,0)$ and the second one interpolates from $n \propto (0,1,0)$ to $n \propto (1,0,0)$.  We also have another possibility, which 
cannot be seen in the $w$ patch, and which corresponds to $n$ being of the form 
\beq
n \propto   \left( \begin{array}{c} 
0\\
1/x_+ \\
\zeta x_+
\end{array} \right) 
\eeq
This expressions corresponds to a third parton which interpolates from $n \propto (0,1,0)$ to $n \propto (0,0,1)$.
All these three partons together cover the the perimeter of the toric diagram in Figure \ref{toric}.

We have thus seen that the conformal map (\ref{conformalmapping}) is a convenient tool to analyse solutions on $\R \times S^1$. In particular, it maps the infinite chain of solitons to a finite number of them related by some form of the  $\Z_h$ symmetry.  Also, the 
fractionalization has a simple interpretation here.  There is no actual fractionalization in the $x_+$ plane, but there can be in the $z$ cylinder case due to the map  (\ref{conformalmapping}) when $h$ is greater than one.

 So far we have just re-derived some known results \cite{Bruckmann:2007zh,Brendel:2009mp,Harland:2009mf}, but for the non-holomorphic solutions this map will be particularly useful to perform the required computations and to derive new results.

\section{Non-holomorphic solutions and their generators}
\label{nonholosection}

Given a generic   $N$  vector  $f_I$   its corresponding  state in $\CP^{N-1}$ in the $n$-formulation is 
\beq
n_I = \frac{f_I}{|f|}.
\eeq
Note that this choice corresponds to (\ref{aa}) in which  $f_I$ was obtained from $w$ by multiplying it by all denominators of the components of $w$ (and then dropping all common factors).

The $P_+$ operator is defined to act on  $f$  as follows:
\beq
\label{defpplus}
(P_+f)_I =  \partial_+ f_I  - \frac{f^{\dagger}_J \partial_+ f_J }{|f|^2}f_I.
\eeq
Thus the $P_+$ operator retains only the part of $\partial_+ f_I$  which is  orthogonal to $f_I$. 
Its related vector $P_+ n$ is given by 
\beq
(P_+ n)_I = \frac{(P_+ f)_I}{|P_+ f|}
\eeq
It is easy to see that $P_+ n$ only changes by an irrelevant overall phase if the vector $f$ is multiplied by an arbitrary function. As is well known\cite{Din:1980jg} the $P_+$ operator maps  a solution of the $\CP^{N-1}$ sigma model into another solution, that is if $n$ solves the Euler-Lagrange equations (\ref{elequation}) then so does $P_+ n$ \cite{Din:1980jg}.

It is convenient to introduce a wedge product formulation.  We define  $f\wedge \partial_+f$  to be
\beq
(f\wedge \partial_+ f)_{IJ}\,=\, f_I \partial_+f_J\,-f_J \partial_+f_I
\eeq
and similarity  for $(f\wedge \partial_+f\wedge \partial_+^2f)_{IJK}$.
In this formulation the $P_+$ operator (\ref{defpplus}) is given by
\beq
(P_+f)_I = \frac{f_{J}^{\dagger}(f\wedge \partial_+f)_{IJ}}{|f|^2}
\eeq
and its repeated action
\beq
(P_+^2f)_I = \frac{(f\wedge \partial_+f)_{JK}^{\dagger}(f\wedge \partial_+f\wedge \partial_+^2f)_{IJK}}{|f\wedge\partial_+f|^2}
\eeq
In the $\CP^2$ case the wedge products simplify considerably. We can introduce a vector $B$ and  a  function $A$ defined as follows
\bea
\label{BAdef}
(f\wedge \partial_+f)_{IJ} &=& \frac{1}{2}\epsilon_{IJK}
B_K, \nonumber \\ 
(f\wedge \partial_+ f \wedge \partial_+^2f)_{IJK} &=& \epsilon_{IJK}A.
\eea

With these definitions we find that, up to overall factors, which cancel in $n$ 
\bea
\label{pab}
(P_+f)_I &\propto&-\frac{\epsilon_{IJK}\,f^{\dagger}_J\,B_K}{\vert f\vert^2}, \nonumber \\
(P_+^2f)_I &\propto& -\frac{B_I^{\dagger}A}{\vert \partial_+f\vert^2}.
\eea

The expressions for the actions of the corresponding solutions are given by
\beq
S[n] = R_1 \qquad S[P_+ n] = R_1 + R_2,
\eeq
where 
\beq
R_1=\frac{\vert P_+f\vert^2}{\vert f\vert^2}\qquad \hbox{and}\qquad R_2=\frac{\vert P_+^2f\vert^2}{\vert P_+f\vert^2}.
\eeq
All this is true also in a general $\CP^{N-1}$ model.  For $\CP^2$ we also have  
$S[P_+^2n] = R_2$ and so 
\beq
\label{sumactions}
S[P_+n] = S[n] + S[P_+^2n]
\eeq

Moreover for the $\CP^2$ case the expressions for $R_{1,2}$  simplify further:
\beq
\label{R12def}
R_1=\frac{\vert B\vert^2}{\vert f\vert^4} \qquad \hbox{and} \qquad 
R_2=\frac{\vert A\vert^2 \vert f\vert^2}{\vert B\vert^4}.
\eeq

As is well known, for $\CP^1$ there are only holomorphic and anti-holomorphic solutions.
If $f=\left( \begin{array}{c} 
1 \\
w
\end{array}\right)$ with $w$ holomorphic then
\beq
P_+f \propto \left( \begin{array}{c} 
-\bar{w} \\
1
\end{array}\right)
\eeq
and so the action of $P_+$ on $k$ solitons creates $k$ anti-solitons and a further action of $P_+$ on the anti-solitons produces 
vacuum. 
We can characterize a generic solution by a pair of integers  $(a,b)$ where $a$ is the number of solitons and $b$ the number of anti-solitons.  In this terminology, the chain of the operations of the $P_+$ operator gives
\beq 
\label{cp1embeddedchain}
(k,0) \xrightarrow{\mbox{\tiny{$P_+$}}}     (0,k)  \xrightarrow{\mbox{\tiny{$P_+$}}}    (0,0). 
\eeq

For $\CP^2$ there are in addition also non-BPS solutions.   A one-soliton solution can always be treated as lying in a $\CP^1$ space embedded into $\CP^2$.  So the action of the $P_+$ operator on such a field configuration does not produce any non-trivial non-BPS solutions. To obtain non-BPS solutions the initial configuration has contain at least two solitons. 

Consider a $\CP^2$ holomorphic solution describing $k$ solitons which in the $w$ patch is given by 
\beq
\left( \begin{array}{c} 
w_1 \\
w_2 \end{array} \right) =  \sum_{\alpha=1}^k\left(\begin{array}{c}
\frac{\zeta_{\alpha}}{x_+-\xi_{\alpha}}\\
\frac{\varrho_{\alpha}}{x_+-\xi_{\alpha}}
\end{array}\right).
\eeq
The complex numbers $\zeta_{\alpha},\varrho_{\alpha},\xi_{\alpha}$ parametrize a moduli space of dimension $6k$.  This is the most generic $k$-soliton solution with $n=(1,0,0)$ fixed at infinity.  To compute the polynomial vector $f$ we first find the corresponding $n$ as defined in (\ref{nfromw}) and then  bring all terms to a common denominator so that the final vector takes the form
\beq
f=\left( \begin{array}{c} 
\prod_{\alpha=1}^k(x_+-\xi_{\alpha})\\
\sum_{\alpha=1}^k \zeta_{\alpha}\prod_{\beta \neq \alpha }^k(x_+-\xi_{\beta})\\
\sum_{\alpha=1}^k \varrho_{\alpha}\prod_{\beta \neq \alpha}^k(x_+-\xi_{\beta})
\end{array}\right).
\eeq
The first component of $f$ has degree $k$ while the others have at most degree $k-1$.
We know from (\ref{pab}) that $P_+^2 f \propto B^{\dagger}$ with $B$ given in (\ref{BAdef}), so $ P_+^2 f$ contains only anti-solitons and their maximal number is $2k-2$.
The chain of  actions of the $P_+$ operator for $\CP^2$ is thus in general 
\beq 
\label{max}
(k,0) \xrightarrow{\mbox{\tiny{$P_+$}}}   (2k-2,k) \xrightarrow{\mbox{\tiny{$P_+$}}}    (0,2k-2) 
\eeq

Note that $2k-2$ is  an upper-bound on the possible number of anti-instantons in the $P_+^2 f$ solution, but  for some configurations the number of anti-solitons that are generated can be smaller.
The simplest example is the case in which all the $k$ solitons can be embedded into a  $\CP^1 \subset \CP^2$ for which (\ref{cp1embeddedchain}) must be true.
The moduli space of $k$ solitons in $\CP^2$ has dimension $6k$ while the moduli space of $k$ solitons embedded in a generic $\CP^1$ has dimension $4k+2$ (the extra $2$ comes from the different embeddings). Moreover,  there are other sub-manifolds where $b$, the dimension of this subspace, satisfies $0<b<2k-2$.

For example, if we impose the conditions
\beq
\label{firstconstraint}
\sum_{\alpha} \zeta_{\alpha}=0 \qquad {\rm and} \qquad   \sum_{\alpha}\varrho_{\alpha}=0
\eeq
then  we have at most $2k-3$ solitons. To have at most $2k-4$ solitons we have to impose also
\beq
\sum_{\alpha\neq \beta} \zeta_{\alpha} \xi_{\beta} =0\qquad {\rm and} \qquad   \sum_{\alpha \neq \beta}\varrho_{\alpha} \xi_{\beta}=0
\eeq
and so on up to a minimum of $k-1$. So we have a sequence of nested manifolds of dimension $6k-4d$ which have at most $2k-2-d$ anti-solitons in $P_+^2f$ with $d=0,1,\dots,k+1$. Note that the  manifold of dimension $4k+2$ of solutions embeddable in $\CP^1$ is not completely contained in the previous ones, as it is clear, for example, by considering the dimensionality of the smallest manifold whose dimension is $2k-4$. The reason for this is that reducing the degree of polynomials in $f$ is not the only way to reduce the number of instantons. This can be achieved also by arranging for all $f_{1,2,3}$ to have a common factor. Such conditions are easy to discuss in each concrete case but much harder to describe for a general configuration.

\section{Examples: embedding,  clustering and de-compactification limits}
\label{example}

The conformal mapping (\ref{conformalmapping}) maps solutions of the Euler-Lagrange equations into  other solutions,  and this is true not only for the holomorphic  or anti-holomorphic solutions but also for the non-BPS ones.   So all the solutions on the cylinder are mapped into the  subset of solutions in the $x_+$ plane which satisfy  a certain $\Z_h$ constraint.    The theorems proved in \cite{Din:1980jg,Din:1980uj,Din:1980wh} which assumed the finiteness of the total action can then be applied too and 
 we also know that the $P_{\pm}$ operators generate all solutions.

To see how this works let us first take two solitons in the $x_+$ plane of size $\lambda$ located at  positions $\pm1$. In the $w$ formulation they are given by
\beq 
\label{example22}
 \left( \begin{array}{c} 
w_1 \\
w_2 \end{array} \right)
=
 \left( \begin{array}{c} 
\frac{ c \lambda  }{ x_+ -1} \\
 \frac{s \lambda }{x_+ - 1}  \end{array} \right) +  
\left( \begin{array}{c} 
\frac{ c \lambda}{ x_+ +  1 } \\
 \frac{-s \lambda }{x_+ +  1}  \end{array} \right),
\eeq
where we have abbreviated $s=\sin{(\alpha)}$, $c =\cos{(\alpha)}$.
Each soliton is separately embedded in a $\CP^1 \subset \CP^2$, but the configuration is not the most general 2 soliton solution of $\CP^2$. The angle $\alpha$ parametrizes the different orientations of these embeddings.  In particular, the whole solution can be embedded in a unique $\CP^1$ only for the special cases $\alpha = 0,\pi/2$.  These solutions describe a $\Z_2$ symmetric configuration corresponding to the boundary condition
\beq
\label{z2symmetry}
\Omega = {\rm diag} (1,-1,1),
\eeq
which can be obtained via the conformal mapping (\ref{conformalmapping})  with $h=2$.   This case is simpler than the one with the twisted boundary conditions  and realizes all the phenomena that are of interest to us. 
The vector $f$, with polynomial components, corresponding to (\ref{example22}) is 
\beq
\label{examplez2}
f= \left( \begin{array}{c} 
x_+^2 - 1 \\
  2 c\lambda\,  x_+ \\
 2 s \lambda \end{array} \right).
\eeq
Computing $B$ and $A$, defined in (\ref{BAdef}), we get 
\beq
B= \left(\begin{array}{c}
-4 sc\lambda^2  \\
4s\lambda x_+\\
-2c\lambda( x_+^2+1)
\end{array}\right)
\qquad {\rm and} \qquad 
A=4c\lambda(x_+^2-1). 
\eeq
The two anti-solitons, at the end of the chain (\ref{max}), by using (\ref{pab}), are given by 
\beq
P_+^2 f \propto  \left( \begin{array}{c} 
 2 s \lambda   \\
-2 s   x_-/c
\\
  x_-^2 + 1
 \end{array} \right) = \left( \begin{array}{c} 
 2 \tilde{s}  \tilde{\lambda}   \\
 2\tilde{c}   \tilde{\lambda}  x_-
\\
  x_-^2 + 1
 \end{array} \right).
\label{bbb}
\eeq

The last equality in (\ref{bbb}) involves rewriting the previous expression in a form similar to (\ref{examplez2}) but in a different $w$-patch and  with a different angle $\tilde{\alpha}$ and of different size $\tilde{\lambda}$. We again abbreviated $\tilde{s}=\sin{(\tilde{\alpha})}$ and $\tilde{c} =\cos{(\tilde{\alpha})}$. The relation between the new angle and size and the old ones is given by 
\beq
\tilde{\alpha} = -\arctan{(\lambda c)}, \qquad \qquad
\tilde{\lambda} = - t \sqrt{1+ c^2 \lambda^2}.
\eeq
So the  two anti-solitons, at the end of the chain, are always located at fixed positions  $\pm i$ and  have their relative orientation and size determined by $\tilde{\alpha}$, $\tilde{\lambda}$.

The first interesting limit to explore is when the relative orientation of the two solitons goes to zero. This limit corresponds to the case (\ref{cp1embeddedchain}) in which the whole holomorphic solution can be embedded into a unique $\CP^1$ and thus no anti-solitons are created. We know that this is the case when $\alpha$ is strictly zero but we want to see how,  when the limit is taken, the two solitons disappear.  Taking the limit corresponds to letting 
\beq 
\label{embeddinglimit}
\begin{array}{c} 
\alpha \to 0    \\
    \lambda \ {\rm fixed}
 \end{array} 
\qquad
\quad
 \Longrightarrow 
\qquad
\quad
\begin{array}{c} 
 \tilde{\alpha} \to -\arctan{(\lambda)} \\
    \tilde{\lambda} \to  0
 \end{array}.
\eeq
Note that as  $\tilde{\lambda} \to 0$ the two solitons, while remaining at fixed positions $\pm i$, become singular (and disappear as `delta' functions).  So the limit is  not continuous from the point of view of the action
\beq
4\pi = \lim_{\alpha \to 0}{S[P_+^2 n(\alpha,\lambda)]}  \neq S[P_+^2 n(0,\lambda)] = 0
\eeq

The other embedding limit corresponds to  $\alpha \to \pi/2$. This time,  the two anti-solitons disappear by becoming infinitely wide
\beq 
\label{embeddinglimitdue}
\begin{array}{c} 
\alpha \to \pi/2    \\
    \lambda \ {\rm fixed}
 \end{array} 
\qquad
\quad
 \Longrightarrow 
\qquad
\quad
\begin{array}{c} 
 \tilde{\alpha} \to 0 \\
    \tilde{\lambda} \to  \infty
 \end{array}.
\eeq

Another interesting limit is that of the clustering, which corresponds to sending the distance between the two solitons to infinity while keeping their sizes and their relative angle fixed.  By conformal invariance the clustering limit  is equivalent to the limit  $\lambda \to 0$ while keeping the distance fixed, and thus we can use the solutions of the form (\ref{example22}). This limit, on the anti-solitons $P_+^2 f$,  has the following effect
\beq
\label{clusteringlimit} 
\begin{array}{c} 
\alpha \  {\rm fixed}   \\
    \lambda \to 0
 \end{array} 
\qquad
\quad
 \Longrightarrow 
\qquad
\quad
\begin{array}{c} 
 \tilde{\alpha} \to 0 \\
  \tilde{\lambda} \to -  t
 \end{array},
\eeq
where $t$ is some number.
Note that the anti-solitons remain of the fixed size in this limit. Note also that this observation demonstrates that the action of the operator $P_+$ is highly non-local and, in particular, the states it produces may not obey the clustering property. The two original solitons, after computing $P_+$, have an effect on each-other which does not decrease or disappear as their distance is sent to infinity.

\begin{figure}[h!]
\centerline{
\epsfxsize=4.5cm\epsfbox{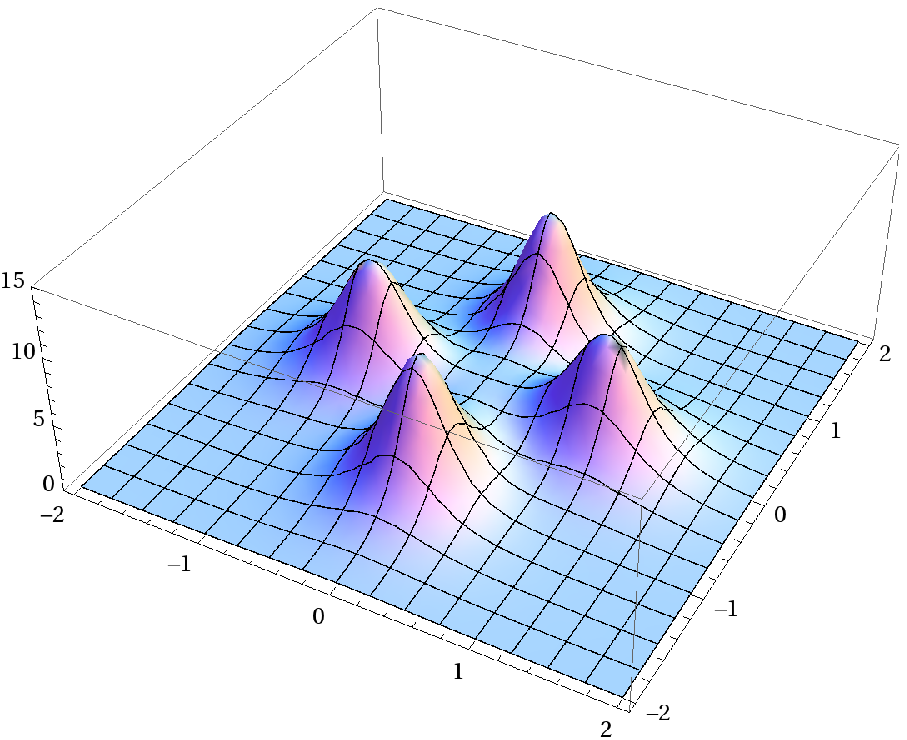} 
\epsfxsize=4.5cm\epsfbox{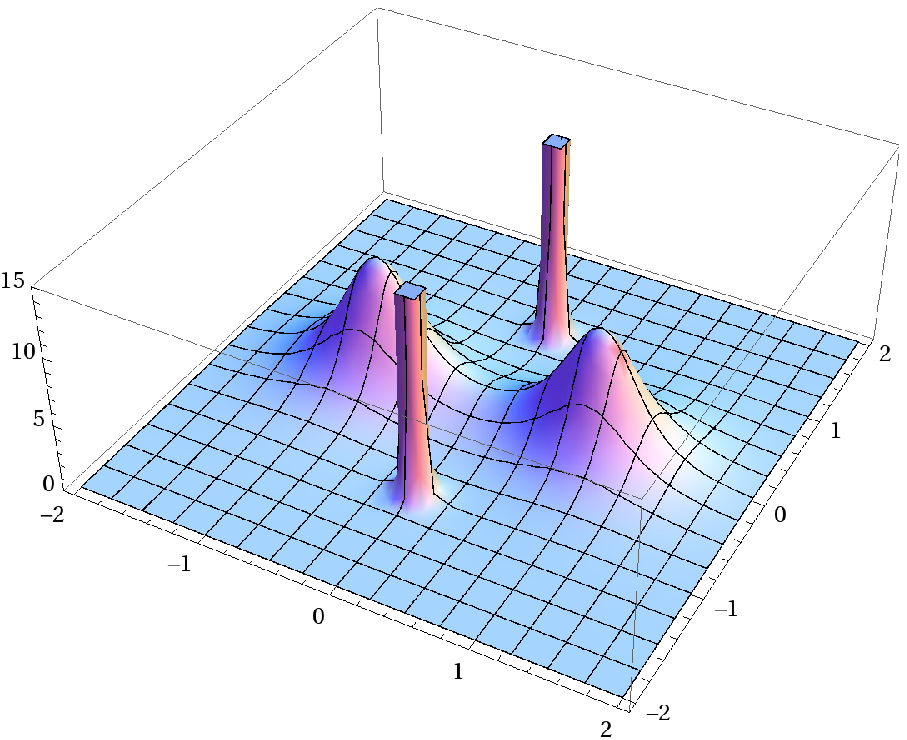}
\epsfxsize=4.5cm \epsfbox{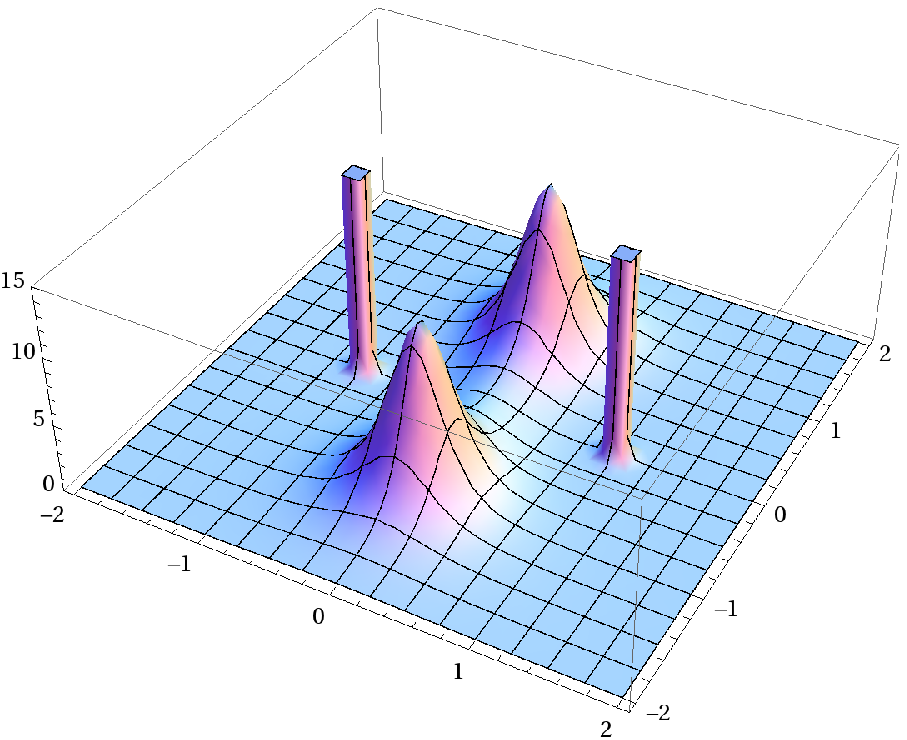}}
\caption{{\footnotesize  Action density for $P_+ f$  with $f$ given by the two-soliton example (\ref{examplez2}). $P_+ f$ is a mixture of two anti-solitons located at $x_+=\pm 1$ and of two solitons located at $x_+=\pm i$. The three plots correspond, respectively, to $(\alpha,\lambda) = (.4,.5), (.01,.5), (.4,.005) $.  The second plot corresponds  to the $\alpha$ small case, thus very close to the embedding limit (\ref{embeddinglimit}).  The third one corresponds  to the $\lambda$ small case  thus very close to the clustering limit (\ref{clusteringlimit}).   }}
\label{twoz2}
\end{figure}

The above mentioned analysis has been done on the final step of the chain, that is  on $P_+^2f$. 
The intermediate step $P_+f$ is, in general, a mixture of the two initial and the two final solitons. So we expect the two limits (\ref{embeddinglimit}) and (\ref{clusteringlimit}) to produce the same result. Some examples of $S[P_+n]$ are given in Figure \ref{twoz2} for different values of $\alpha$ and $\lambda$. 
The tools described in Section \ref{nonholosection} allow us to confirm this also analytically. 
Computing $R_1$ and $R_2$ defined in (\ref{R12def}) for the solution (\ref{examplez2}) gives us
\bea
R_1 &=& \frac{4 \lambda^2 \left( 4 s^2c^2\lambda^2+4s^2\vert x_+\vert^2+c^2\vert x_+^2+1\vert^2\right)}{(4s^2\lambda^2+
4c^2\lambda^2\vert x_+\vert^2+\vert x_+^2-1\vert^2)^2},
\nonumber \\
R_2 &=& 
\frac{4c^2s^2\lambda^4(\vert x_+^2-1\vert^2+4c^2\lambda^2\vert x_+\vert^2+s^2\lambda^2)}{(4s^2c^2\lambda^4+4s^2\lambda^2\vert x_+\vert^2+c^2\lambda^2\vert x_+^2+1\vert^2)^2}.
\eea

The action $S[P_+n]$ is the sum of these two terms, and thus a sum of the action of the initial solution $R_1$ and of the final one $R_2$. For example, in the clustering limit $\lambda \to 0$,
we have 
\bea
R_1 &\sim&
\frac{4 \lambda^2 \left( 4 s^2\vert x_+\vert^2+c^2\vert x_+^2+1\vert^2\right)}{(4s^2\lambda^2+4
c^2\lambda^2\vert x_+\vert^2+\vert x_+^2-1\vert^2)^2},
\nonumber \\
 R_2 &\rightarrow& \frac{4c^2s^2\vert x_+^2-1\vert^2}{(4s^2\vert x_+\vert^2+c^2\vert x_+^2+1\vert^2)^2}.
\eea
The first term $R_1$ corresponds to the original solitons  located at $\pm 1$ shrinking to zero  while the second term describes the other two  solitons located at $\pm i$ which remain of fixed size and which do not disappear.

Next we move to the twisted boundary conditions (\ref{twistedconditioncp2}). A one-soliton solution in the $z$ cylinder corresponds in the $x_+$ plane to the following three-solitons with  $\Z_3$ symmetry
\beq 
\label{threez3}
\left( \begin{array}{c} 
w_1 \\
w_2 \end{array} \right) = 
 \left( \begin{array}{c} 
\frac{ c \lambda }{ x_+ -1 } \\
 \frac{s \lambda}{x_+ - 1}  \end{array}
 \right) 
+ 
\left( \begin{array}{c} 
\frac{\omega^2   c   \lambda }{ x_+ -  \omega } \\
\frac{\omega  s   \lambda }{ x_+ -  \omega}  \end{array} \right) 
+ 
\left( \begin{array}{c} 
\frac{\omega     c   \lambda }{ x_+ -  \omega^2 } \\
\frac{\omega^2   s   \lambda }{ x_+ -  \omega^2}  \end{array} \right), 
\eeq
where $\alpha$, as before, is an angle parametrizing the $\CP^1$ of each soliton embedded into $\CP^2$. The polynomial vector $f$ and $P_+^2 f$ for this solution  are given by 
\beq
\label{examplez3}
f= \left( \begin{array}{c} 
x_+^3 - 1 \\
 3 c\lambda x_+ \\
 3 s \lambda \end{array} \right)
\qquad \Rightarrow \qquad 
P_+^2 f \propto  \left( \begin{array}{c} 
 3 \lambda  s \\
  -3  x_-^2 t \\
1+2 x_-^3\end{array} \right).
\eeq

\begin{figure}[h!]
\centerline{
\epsfxsize=4.5cm\epsfbox{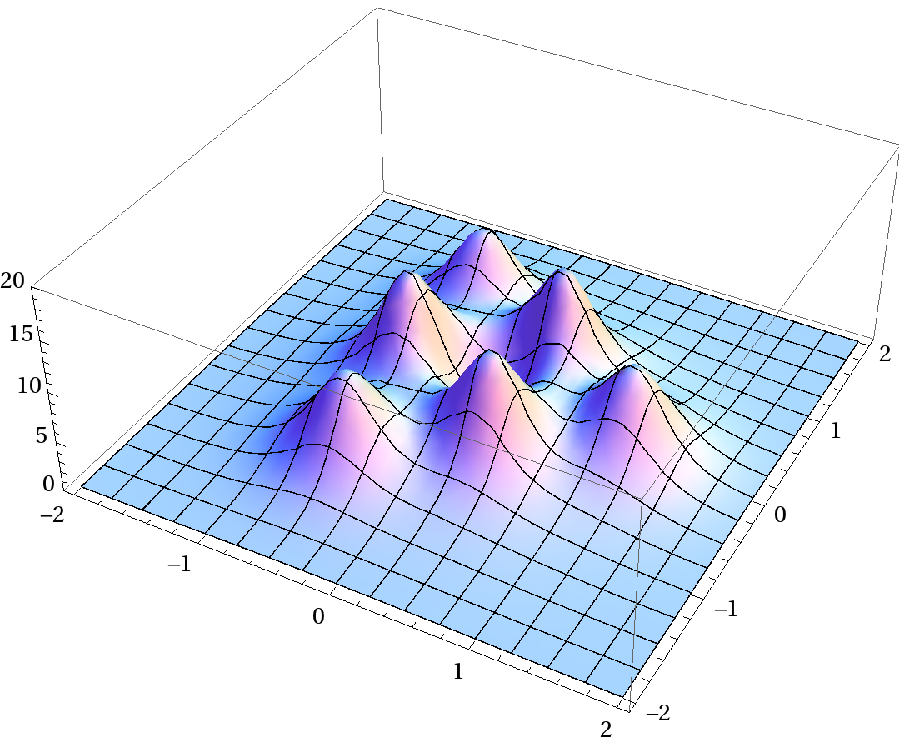}
\epsfxsize=4.5cm \epsfbox{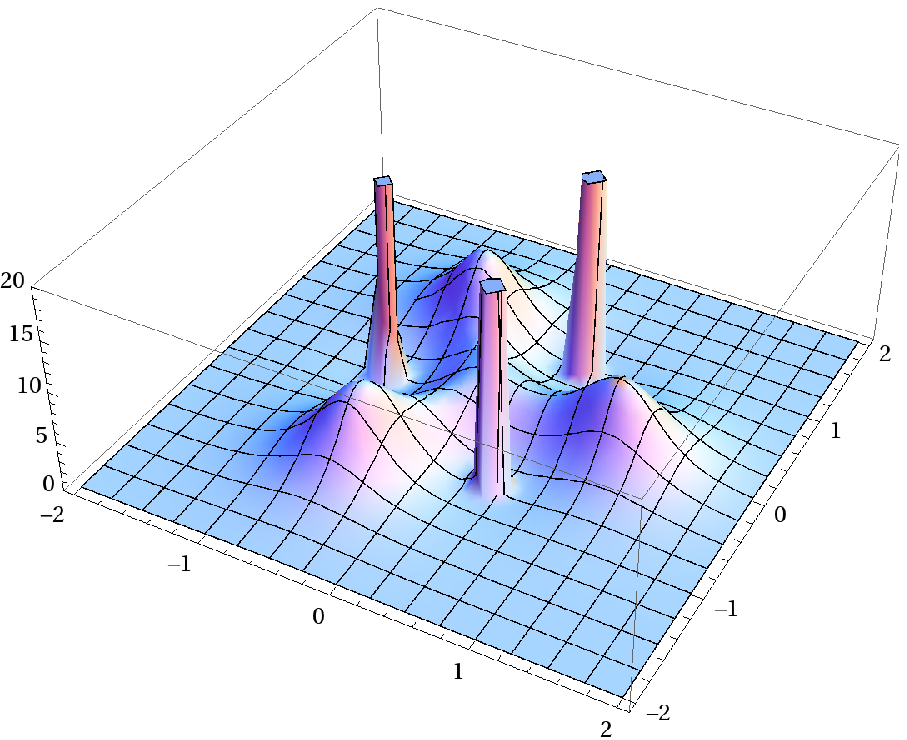}
\epsfxsize=4.5cm \epsfbox{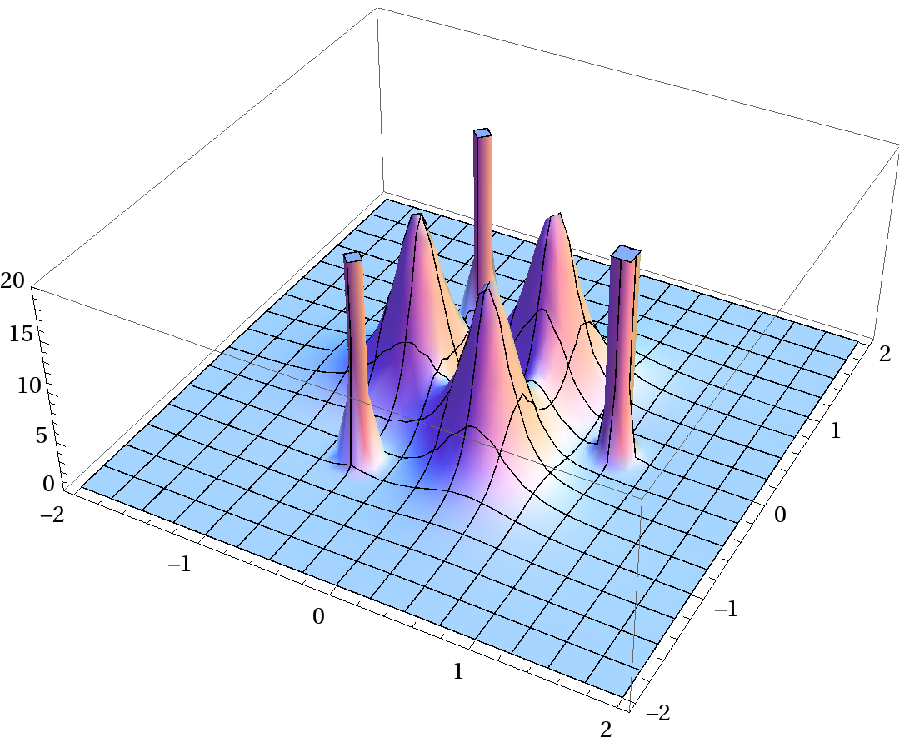}}
\caption{{\footnotesize  Action density for $P_+ f$  for $f$ given by the three-soliton example in (\ref{examplez3}). $P_+ f$ is a mixture of three  anti-solitons located at $1,\omega, \omega^2$ and three solitons located at $\rho,\rho \omega, \rho \omega^2$. The three plots correspond respectively to $(\alpha,\lambda) = (.8,.5),\, (.015,.5),\, (.8,.01) $. The second plot corresponds  to the $\alpha$ small case; the third one   to the $\lambda$ small case. }}
\label{threez3fig}
\end{figure}

 To interpret this result we need to decompose $P_+^2 f$ into its anti-soliton components. For this we need to use a different $w$-patch in which
\beq
\label{nfromwnewpatch}
n = \frac{1}{\sqrt{1+|\tilde{w}|^2}}
 \left( \begin{array}{c} 
 \tilde{w}_1 \\
\tilde{w}_2\\
1 \end{array} \right)
\eeq  
as then $P_+^2 f$ can be decomposed into the following three anti-solitons
\beq 
\label{threez3sym}
\left( \begin{array}{c} 
\tilde{w}_1 \\
\tilde{w}_2 \end{array} \right) = 
 \left( \begin{array}{c} 
\frac{\omega^2 \tc \tlambda }{ x_- -  \rho } \\
 \frac{\ts \tlambda}{x_- - \rho}  \end{array}
 \right) 
+ 
\left( \begin{array}{c} 
\frac{  \tc   \tlambda }{ x_- -  \omega \rho } \\
\frac{   \ts   \tlambda }{ x_- -  \omega \rho}  \end{array} \right) 
+ 
\left( \begin{array}{c} 
\frac{  \omega   \tc   \tlambda }{ x_- -  \omega^{2} \rho } \\
\frac{   \ts   \tlambda }{ x_- -  \omega^{2} \rho}  \end{array} \right), 
\eeq
where $ \rho = \omega^{1/2}/2^{1/3}$, so that
\beq
P_+^2 f \propto  \left( \begin{array}{c} 
3  \tlambda \tc/|\rho|\\
6 x_-^2 \tlambda \ts \\
1+2x_-^3
 \end{array} \right)
\eeq
The positions of the anti-solitons are fixed and the new angle $\tilde{\alpha}$ and size $\tilde{\lambda}$ are given by 
\beq
\tilde{\alpha} = - \arctan{\left(\frac{|\rho|^2}{\lambda c}\right)}, \qquad \qquad
\tilde{\lambda} = \frac{t}{2}  \sqrt{1 + \frac{\lambda^2 c^2}{|\rho|^4} }
\eeq
and we see that 
$\tlambda \to 0$ as $\alpha \to 0$ and 
$\tlambda \to t/2$ as $\lambda \to 0$.
Note that the applications of the $P_+$ operator generates the following chain
\beq 
(3,0) \xrightarrow{\mbox{\tiny{$P_+$}}}   (3,3) \xrightarrow{\mbox{\tiny{$P_+$}}}    (0,3) 
\eeq
for which the soliton numbers are smaller than in (\ref{max}) since (\ref{threez3}) satisfy the constraint (\ref{firstconstraint}).  The intermediate configuration $P_+ f$ is a mixture containing six solitons in total.
The embedding limit ($\alpha \to 0$ with $\lambda$ fixed) and the clustering limit ($\lambda \to 0$ with $\alpha$ fixed)  have the same features as in the previous example; some examples as shown in Figure \ref{threez3fig}. 
When $\alpha \to 0, \pi/2$ the anti-solitons disappear by becoming very peaked or very spread out. In the clustering limit $\lambda \to 0$ the anti-solitons remain of fixed size.

Dabrowski and Dunne in \cite{Dabrowski:2013kba} considered some fractionalized non-BPS solutions on $\R \times S^1$, and the simplest example is given by $P_+$ acting on two partons with different orientations inside $\CP^2$.
In the $x_+$ plan this corresponds to the solution
\beq
\left( \begin{array}{c} 
w_1 \\
w_2 \end{array} \right)
=
 \left( \begin{array}{c} 
 \frac{\zeta}{ x_+}  \\
 \frac{\xi}{ x_+^2}   \end{array} \right), 
\eeq
which is the most general two-instanton solution which satisfies the $\Z_3$ constraint (\ref{threez3sym}) and has $n \propto (1,0,0)$ at infinity.
For $\zeta=0$ we have two coincident axial symmetric solitons embedded in the same $\CP^1$ and for $\zeta \neq 0$ they have different orientations. 
 We can then use the scale invariance to fix, for example, $\xi =1$ and this fixes the centre of mass of the two partons to be, in the cylinder coordinates, at $\Im{(z)}=0$.  The other parameter $\zeta$ represents the distance, which is  $d= 2 \log{|\zeta|}/\pi $ for large $|\zeta|$, and the relative phase between the two partons.  Following the same procedure as in  the previous examples, we find $f$ and $P_+^2f$ to be given by 
\beq
f = \left( \begin{array}{c}  
 x_+^2\\
\zeta x_+\\
\xi\\
\end{array} \right) 
\qquad \Rightarrow \qquad 
P_+^2 f \propto  \left( \begin{array}{c}  
 \xi^*  \\
 - 2 \xi^* x_- / \zeta^* \\
 x_-^2\\
\end{array} \right), 
\eeq
which thus tells us that at the end of the chain we have two anti-solitons. The $P_+ n$ is thus a $(2,2)$ soliton solution which in the $z$ cylinder corresponds to 4 total partons. The  $P_+^2 f$  is very similar to $f$ so it can be interpreted in the same way but in a different patch and with
\beq
\tilde{\xi} =\xi^* \qquad \qquad \tilde{\zeta} = - \frac{2\xi^*}{\zeta^*}
\eeq
Note that as $\zeta \to 0$ we have $\tilde{\zeta}\to \infty $ so in the embedding limit  the two anti-solitons disappear by going to infinite distance in $\Im{(z)}$.

Finally we consider a $4$ soliton configuration. This case will allow us to demonstrate an even more severe pathology of the clustering decomposition. This time we consider 4 solitons  with the $\Z_2$ symmetry (\ref{z2symmetry})
\beq
\label{examplez24} \left( \begin{array}{c} 
w_1 \\
w_2 \end{array} \right)
=
 \left( \begin{array}{c} 
\frac{ c \lambda }{ x_+ -  a} \\
 \frac{s \lambda}{x_+ - a}  \end{array} \right) + 
\left( \begin{array}{c} 
\frac{ c \lambda }{ x_+ -b } \\
 \frac{s \lambda}{x_+ - b}  \end{array} \right)
+ 
\left( \begin{array}{c} 
\frac{ c \lambda }{x_+ +  a} \\
 \frac{-s \lambda}{x_+ + a}  \end{array} \right)
+ 
\left( \begin{array}{c} 
\frac{ c \lambda }{ x_+ + b} \\
 \frac{-s \lambda}{x_+ +  b}  \end{array} \right).
\eeq

This configuration consists of two solitons located at $a$ and $b$ and embedded in the same $\CP^1$ together with their $\Z_2$ symmetric partners located at $-a$ and $-b$. So all together we have two clusters each composed of two solitons which are separately embeddable into $\CP^1$. The relative angle $\alpha$, as before,  parametrizes the relative embedding of the two clusters. The polynomial vector $f$ is now
\beq
f= \left( \begin{array}{c} 
 (x_+^2 - a^2)(x_+^2 - b^2) \\
 2 c \lambda  x_+ (2x_+^2 - a^2-b^2)\\
2 s \lambda (a+b) (x_+^2 - ab) \end{array} \right).
\eeq
\begin{figure}[h!]
\centerline{
\epsfxsize=4.5cm\epsfbox{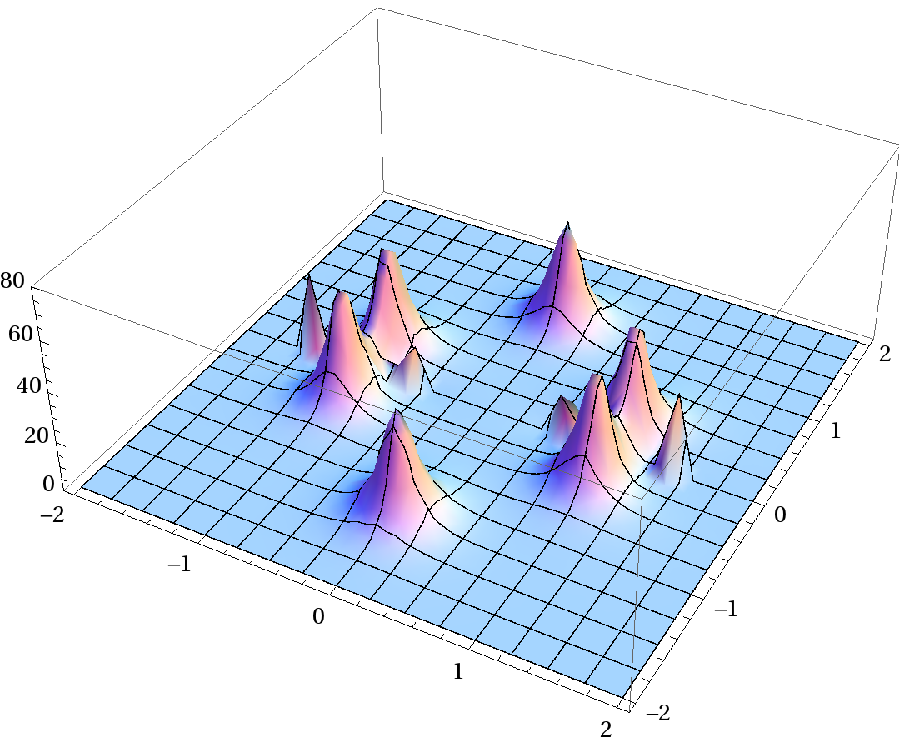}
\epsfxsize=4.5cm \epsfbox{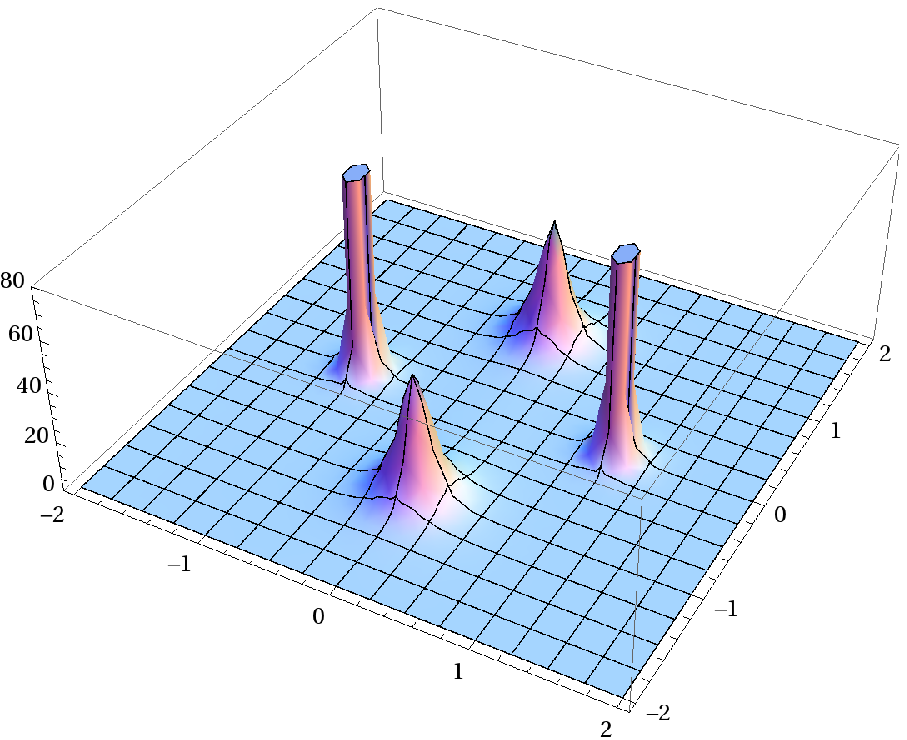}}
\centerline{
\epsfxsize=4.5cm\epsfbox{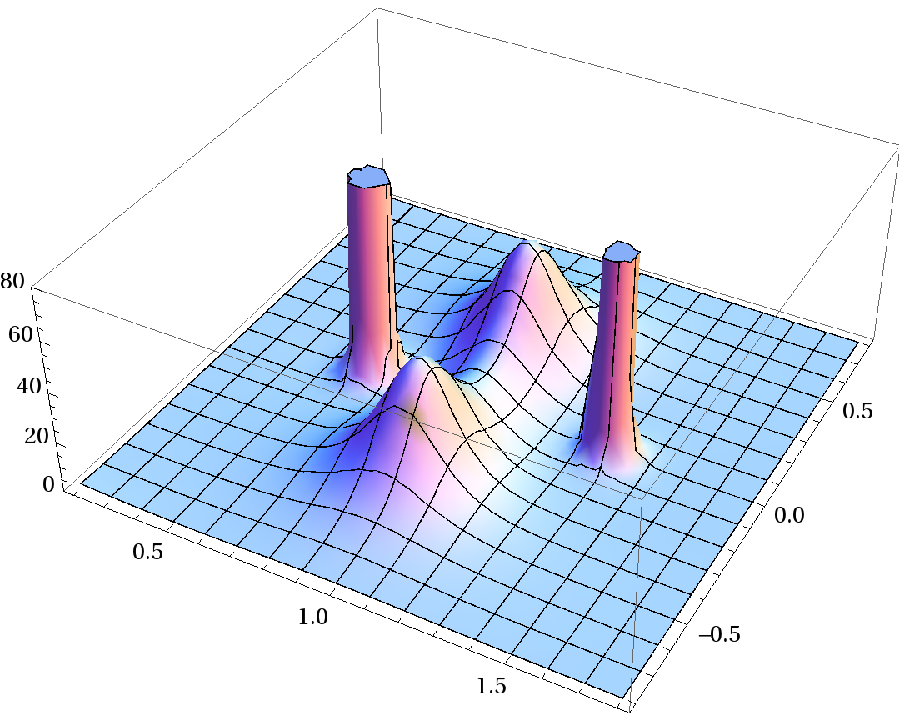}
\epsfxsize=4.5cm \epsfbox{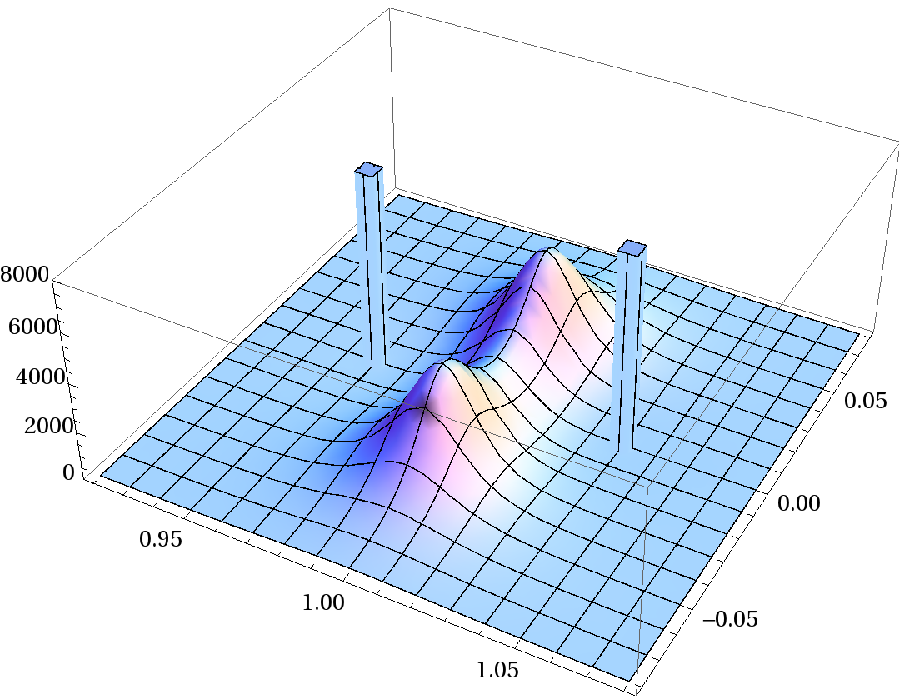}}
\caption{{\footnotesize The first line is the action density for $P_+ f$ of the $\Z_2$ symmetric configuration with $4$  anti-solitons plus $6$ solitons (\ref{examplez24}) for the values $(\alpha,\lambda) = (.2,.25),\, (.2,.025) $ and $\epsilon = 1.6 \lambda$.  The second line is obtained by zooming into the right-hand cluster for the same values of $(\alpha,\lambda)$. }}
\label{fourz2}
\end{figure}

To study the clustering limit we consider
\bea
\label{cluster42}
a=1+i \epsilon \quad {\rm and} \quad  b = 1-i \epsilon \nonumber \\
 \lambda, \epsilon \to 0 \qquad  {\rm keeping} \quad  \frac{\lambda}{\epsilon},\ \alpha=\ {\rm fixed}
\eea
This is equivalent to sending the two clusters to an infinite distance from each other. In this limit  two of the anti-solitons remain at  a fixed position  and have a finite limiting size as before (see first line of Figure \ref{fourz2}).
To detect the other four anti-solitons we have to zoom into the individual clusters (see second line of Figure  \ref{fourz2}).  In this case we follow one cluster by zooming the lengths so that they do not change as $\lambda \to 0$. The other four anti-solitons are individually clustering  with the four solitons and then follow them to the same positions, but shrinking to zero size (becoming $\delta$ functions).

The discontinuities we have found are intrinsic properties of the non BPS solutions and not only of the $P_+$ operator. 
To see this very clearly  let us  take a generalized version of (\ref{examplez24}) given by 
\bea
\label{generalizationz24}
\left( \begin{array}{c} 
w_1 \\
w_2 \end{array} \right)
&=&  \left( \begin{array}{c} 
\frac{e^{i \phi} \cos{(\alpha + \beta)} \lambda }{ x_+ -  a} \\
 \frac{e^{-i \phi}\sin{(\alpha + \beta)} \lambda}{x_+ - a}  \end{array} \right) + 
\left( \begin{array}{c} 
\frac{e^{i \phi} \cos{(\alpha - \beta)} \lambda }{ x_+ -b } \\
 \frac{e^{-i \phi}\sin{(\alpha - \beta)} \lambda}{x_+ - b}  \end{array} \right)\nonumber \\
&&
+ 
\left( \begin{array}{c} 
\frac{ e^{i \phi} \cos{(\alpha + \beta)}\lambda }{x_+ +  a} \\
 \frac{-e^{-i \phi}\sin{(\alpha + \beta)} \lambda}{x_+ + a}  \end{array} \right)
+ 
\left( \begin{array}{c} 
\frac{e^{i \phi} \cos{(\alpha - \beta)} \lambda }{ x_+ + b} \\
 \frac{-e^{-i \phi}\sin{(\alpha - \beta)} \lambda}{x_+ +  b}  \end{array} \right) 
\eea
The angles $\alpha$ and $\phi$ describes the relative orientation between the two clusters inside $\CP^2$,  while the angle $\beta$ describes the relative orientation between the two solitons within each cluster.  We want to consider the clustering limit  of $P_+ f$ by using (\ref{cluster42}) and  keeping $\beta$ fixed and, in general, being different from $0,\pi/2$.
In this limit the two clusters flow separately to two non BPS solutions of $2+2$ total solitons each.
By computing explicitly $P_+^2 f$ we can see that there are now six anti-solitons,  in general, and that the number of very thin ones (corresponding to $\delta$ functions) can, at most, be reduced to five by the choice of the relative orientation $\alpha = \pi/2$; thus this number can never drop to four which would be required to have good clustering properties.

Having studied the clustering properties for a number of examples of non-BPS solutions, we can now turn our attention to the same problem for  the $\CP^{N-1}$ model on $\R \times S^1$ (with twisted boundary conditions), and study the correspondence between the solutions at large and small compactification radii in this model. We have already said that, by conformal mapping, the planar limit (or de-compactification limit) of this model  is equivalent to what we called before - the clustering limit in the $x_+$ plane.  As the BPS solutions are continuously connected, the dimension of their moduli space and of the action remain the same during the change of the compactification radius.
A number of discontinuities, however, arises for the non-BPS solutions.  The first class of them is exemplified by the clustering limit of the examples (\ref{example22}) and (\ref{threez3}). These solutions, when pulled back to the $z$ cylinder, do not flow to a localized solution in $\R^2$. There is always a residual contribution on the opposite side of the cylinder coordinates, even when its radius is sent to infinity. This shows that the set of solutions on the cylinder is slightly `larger' than the set of solutions in the plane; there is no one-to-one correspondence between the two formulations.

The example (\ref{generalizationz24}) demonstrates another kind of discontinuity. In this case, after the mapping to the $z$ cylinder, we have a local non-BPS solution which can never be recovered smoothly from a non-BPS solution on the cylinder.
So not only there are more solutions on $\R \times S^1$  than  on $\R^2$, but also there are not, in general,  continuous mappings between them. 

\section{Conclusion}

In this paper we have discussed and compared the solutions of the $\CP^{N-1}$ non-linear sigma model on the plane and on the cylinder. We used a conformal mapping technique to re-derive some results concerning the holomorphic solutions on $\R \times S^1$ with twisted boundary conditions. We then discussed some aspects of the generators of non-holomorphic solutions and showed some discontinuities in these equations that arise in various limits. 
In particular, the examples we studied in Section \ref{example} show some pathological features of the clustering limit for non-BPS solutions due to the non-locality of the $P_+$ operator. Thus if we take  $k$ solitons, divide them into two groups/clusters $k_1 + k_2 =k$ and send these two clusters to infinite distances while keeping fixed their internal structure,  the naive expectation would be that their mutual influence vanishes in this limit.  However, for the $P_+$ operator this is not true, in other words $P_+$ of the $k$ solitons is not equal to $P_+$ of the two individual clusters, even when we send these clusters to infinite distances. 

This also implies that the de-compactification limit of $\R \times S^1$  for the non-BPS solitons is not in general continuous for generic non-BPS solutions.   So the set of solutions of the $\CP^{N-1}$ sigma model on the plane $\R^2$ cannot be continuously related to the set of solutions on the cylinder $\R \times S^1$. 
Any extrapolation of results from small to big radii for the non-BPS solutions must carefully address these issues. 

 Of course, since these discontinuities arise only in the non-BPS sector, they would not affect the one-soliton sector in any  $\CP^{N-1}$  model, which is always BPS. So for example it would not affect the results discussed in \cite{Dunne:2012ae,Dunne:2012zk,Argyres:2012ka} which concerns the first contribution to the trans-series due to one single parton. But to be able to use the compactification method to compute the whole series, or to even prove the existence of a well defined series,  a more detailed analysis of these discontinuities for non-BPS solutions would be required. Note that here we are using the concept of continuity in the strong sense with the natural topology being provided by the action. Trying to define convergence in a weaker sense, if properly formulated, may help in further developments. This problem is currently under active consideration.


\section*{Acknowledgments}
This work is partially funded by the EPSRC grant EP/K003453/1.


\begin{thebibliography}{11}


  

\bibitem{D'Adda:1978uc}
  A.~D'Adda, M.~Luscher and P.~Di Vecchia,
  ``A 1/n Expandable Series of Nonlinear Sigma Models with Instantons,''
  Nucl.\ Phys.\ B {\bf 146} (1978) 63.


\bibitem{Din:1980jg}
  A.~M.~Din and W.~J.~Zakrzewski,
  ``General Classical Solutions in the {CP}$^{(n-1)}$ Model,''
  Nucl.\ Phys.\ B {\bf 174} (1980) 397.

\bibitem{Din:1980uj}
  A.~M.~Din and W.~J.~Zakrzewski,
  ``Properties of General Classical {CP}$^{(n-1)}$ Solutions,''
  Phys.\ Lett.\ B {\bf 95} (1980) 419.

\bibitem{Din:1980wh}
  A.~M.~Din and W.~J.~Zakrzewski,
  ``Interpretation and Further Properties of General Classical {CP}$^{(n-1)}$ Solutions,''
  Nucl.\ Phys.\ B {\bf 182} (1981) 151.

\bibitem{book}
  W.~J.~Zakrzewski,
  ``Low Dimensional Sigma Models,'', Adam Hilger, 1989.


\bibitem{Bruckmann:2007zh}
  F.~Bruckmann,
  ``Instanton constituents in the O(3) model at finite temperature,''
  Phys.\ Rev.\ Lett.\  {\bf 100} (2008) 051602
  [arXiv:0707.0775 [hep-th]].

\bibitem{Brendel:2009mp}
  W.~Brendel, F.~Bruckmann, L.~Janssen, A.~Wipf and C.~Wozar,
  ``Instanton constituents and fermionic zero modes in twisted CP**n models,''
  Phys.\ Lett.\ B {\bf 676} (2009) 116
  [arXiv:0902.2328 [hep-th]].


\bibitem{Harland:2009mf}
  D.~Harland,
  ``Kinks, chains, and loop groups in the CP**n sigma models,''
  J.\ Math.\ Phys.\  {\bf 50} (2009) 122902
  [arXiv:0902.2303 [hep-th]].


\bibitem{Dabrowski:2013kba}
  R.~Dabrowski and G.~V.~Dunne,
  ``Fractionalized Non-Self-Dual Solutions in the CP(N-1) Model,''
  Phys.\ Rev.\ D {\bf 88} (2013) 025020
  [arXiv:1306.0921 [hep-th]].


\bibitem{Cherman:2013yfa}
  A.~Cherman, D.~Dorigoni, G.~V.~Dunne and M.~Unsal,
  ``Resurgence in QFT: Unitons, Fractons and Renormalons in the Principal Chiral Model,''
  arXiv:1308.0127 [hep-th].


\bibitem{Basar:2013eka} 
  G.~Basar, G.~V.~Dunne and M.~Unsal,
  ``Resurgence theory, ghost-instantons, and analytic continuation of path integrals,''
  JHEP {\bf 1310}, 041 (2013)
  [arXiv:1308.1108 [hep-th]].



\bibitem{Argyres:2012ka}
  P.~C.~Argyres and M.~Unsal,
  ``The semi-classical expansion and resurgence in gauge theories: new perturbative, instanton, bion, and renormalon effects,''
  JHEP {\bf 1208} (2012) 063
  [arXiv:1206.1890 [hep-th]].

\bibitem{Dunne:2012ae}
  G.~V.~Dunne and M.~Unsal,
  ``Resurgence and Trans-series in Quantum Field Theory: The CP(N-1) Model,''
  JHEP {\bf 1211} (2012) 170
  [arXiv:1210.2423 [hep-th]].


\bibitem{Dunne:2012zk} 
  G.~V.~Dunne and M.~Unsal,
  ``Continuity and Resurgence: towards a continuum definition of the CP(N-1) model,''
  Phys.\ Rev.\ D {\bf 87}, 025015 (2013)
  [arXiv:1210.3646 [hep-th]].



\end{thebibliography}
\end{document}